\documentclass[doublecol]{epl2}
\usepackage{amsmath}
\usepackage{amssymb}
\usepackage{graphics}
\usepackage{epsfig}

\makeatletter
\def\rightharpoonupfill@{\arrowfill@\relbar\relbar\rightharpoonup}
\newcommand{\overrightharpoonup}{\mathpalette{\overarrow@\rightharpoonupfill@}}

\title{Raising $T_{c}$ in charge density wave superconductor ZrTe$_{3}$ by
Ni intercalation}
\shorttitle{Raising $T_{c}$ in charge density wave superconductor ZrTe$_{3}$ by
Ni intercalation} 

\author{Hechang Lei, Xiangde Zhu,$^{(\ast)}$ and C. Petrovic\thanks{E-mail: \email{petrovic@bnl.gov}}}
\shortauthor{Hechang Lei\etal}

\institute{Condensed Matter Physics and Materials Science Department, Brookhaven
National Laboratory, Upton, NY 11973, USA}


\abstract{We report discovery of bulk superconductivity in Ni$_{0.05}$ZrTe$_{3}$ at $T%
_{c}$ = 3.1 K, obtained through Ni intercalation. Superconductivity coexists
with charge density wave (CDW) state with $T_{CDW}$ = 41 K. When compared to
parent material ZrTe$_{3}$, filamentary superconducting transition is
substantially increased whereas T$_{CDW}$ was suppressed. The analysis of
superconducting state indicates that Ni$_{0.05}$ZrTe$_{3}$ is an
intermediately coupled superconductor.}

\begin{document}

\maketitle

\section{Introduction}

Charge density wave (CDW) and superconductivity are distinctive quantum
orders that emerge due to Fermi surface (FS) instabilities\cite{Gabovich}.
Competition and coexistence between CDW and superconductivity in low-dimensional materials is one of the most fundamental problems in
condensed matter physics. Among CDW\ superconductors, layered MX$_{2}$
compounds and chain-typed MX$_{3}$ compounds (M = transition metal, and X =
S, Se or Te) are well-known\cite{Wilson}$-$\cite{Morosan}. It was recently
pointed out that chalcogenide superconductors represent a weak coupling
limit of the melting of the stripe (or smectic)\ order in cuprates\cite{Sun}%
. Furthermore, the domelike T$_{c}$(x)\cite{Morosan} in Cu$_{x}$TiSe$_{2}$
and pairing mechanism was proposed to arise from the quantum criticality
related to fluctuations in CDW order\cite{Barath}$-$\cite{Kusmartseva}.
Therefore discovery of new superconductors in this materials class is of
significant interest.

ZrTe$_{3}$ belongs to a family of trichalcogenides MX$_{3}$ (M = Ti, Zr, Hf,
U, Th, and X = S, Se, Te). The structure consists of infinite X-X chains
formed by stacking MX$_{3}$ prisms along the crystallographic $\widehat{b}$
axis. The polyhedra are arranged in double sheets and stacked along
monoclinic c axis by van der Waals forces (fig. 1(a))\cite{Furuseth}. ZrTe$%
_{3}$ exhibits CDW transition at $T_{CDW}$ = 63 K. The CDW nesting vector $%
\mathpalette{\overarrow@\rightharpoonupfill@}{q}_{CDW}$ = (1/14, 0, 1/3) has
no component in the chain-axis direction\cite{Eaglesham}. This is different
from another well-known quasi-one-dimensional CDW compound NbSe$_{3}$ with
X-X chains which exhibits a resistivity anomaly along the chain axis\cite%
{Monceau}. The $\rho $(T) of ZrTe$_{3}$ is quasi two dimensional, metallic
below 300 K with anomalies due to CDW transition\cite{Takahashi}, and
superconductivity below $T_{c}$ = 2 K. The superconductivity is not bulk but
filamentary\cite{Nakajima}. Theoretical calculations\cite{Felser} and
photoemission study\cite{Yokoya} indicate that the CDW transition is driven
by the nesting of parallel planar sections of the FS which are related to
the Te 5p$_{x}$ band along the Te-Te chains. Other parts of the FS remain
unaffected and are responsible for superconductivity. Therefore, CDW and
filamentary superconductivity\ coexist in ZrTe$_{3}$ due to multiband nature
of the Fermi surface.

On the other hand, ZrTe$_{3}$ (fig. 1(a)) exhibits uncommon pressure
dependence of CDW and superconductivity\cite{Yomo}. With increasing pressure
$T_{CDW}$ initially increases, then decreases up to 2 GPa and abruptly
vanishes near 5 GPa. The $T_{c}$ falls to below 1.2 K at 0.5 GPa and can not
be observed up to 5 GPa. At higher pressures superconducting transition
emerges again and increases continuously from 2.5 K to 4.7 K at 11 GPa. The $T_{CDW}$(P) and $T_{c}$(P) above 5 GPa can be explained by pressure induced
FS modifications. High pressures favor three-dimensionality, therefore the
area of the planar portions of FS will decrease and the CDW transition will
be suppressed. Consequently superconducting $T_{c}$, a competing FS
instability will be favored. However, the pressure dependence above 2 Gpa is
still unclear. The $T_{c}$(P) and $T_{CDW}$(P) indicate that the changes in
FS are rather sensitive to crystal structure modification. Besides
hydrostatic pressure, doping and intercalation can be used to tune the
crystal structure, lattice properties, shape and characteristics of the FS.
For example, intercalating external atoms into the interlayer weak
van der Waals gap of MX$_{2}$ can depress the CDW transition and enhance the
superconductivity\cite{Morosan}. Since there is also a van der Waals gap in
ZrTe$_{3}$, intercalation is a powerful method to examine the relation
between CDW and superconductivity in this system. Ni is often a
suitable intercalant as seen on the example of Ni$_{x}$TaS$_{2}$
and Ni$_{x}$TaSe$_{2}$\cite{Zhu},\cite{Li2}.

In this work, we report synthesis and physical properties of Ni$_{0.05}$ZrTe$%
_{3}$ single crystals. Ni intercalation induces bulk superconductivity with enhanced $T_{c}$ while the $T_{CDW}$
shifts to lower temperatures, indicating coexistence of CDW and superconductivity.

\section{Experimental}

Single crystal Ni$_{x}$ZrTe$_{3}$ was grown via the chemical vapor transport
(CVT) method. The source and growth zone were set at 700 $^{\circ }C$ for 2
days and then kept at 720 $^{\circ }C$ and 645 $^{\circ }C$ respectively for
10 days. Golden plate-like single crystals with a typical size of 2$\times $2%
$\times $0.2 mm$^{3}$ were obtained. The crystal structure and phase purity
were examined by powder and single crystal X-ray diffraction pattern (XRD)
with Cu K$_{\alpha }$ radiation ($\lambda $ = $1.5418$ \AA ) using a Rigaku
Miniflex X-ray machine. The structure parameters are extracted by fitting
the XRD spectra using the Rietica software\cite{Hunter}. The composition of
Ni$_{x}$ZrTe$_{3}$ single crystal was determined by examination of multiple
points on the crystals using energy dispersive X-ray spectroscopy (EDS) in
an JEOL JSM-6500 scanning electron microscope. The measured compositions are
Ni$_{0.052(3)}$Zr$_{1.00(2)}$Te$_{3.10(2)}$, abbreviated as Ni$_{0.05}$ZrTe$%
_{3}$. Physical property measurements were performed in a Quantum Design
PPMS-9 and MPMS XL 5 instruments.

\section{Results and discussion}

Powder X-ray diffraction (XRD) result (fig. 1(b)) of ground samples
indicates that all peaks of Ni$_{0.05}$ZrTe$_{3}$ can be indexed using the
structure of ZrTe$_{3}$, i.e. the intercalation does not change the
structure of the mother compound. The fitted lattice parameters are a =
0.5895(1) nm, b = 0.3923(1) nm, c = 1.0160(2) nm, and $\beta $ = 97.76(1)$%
^{\circ }$. Our refinement of pure ZrTe$_{3}$ crystals gave lattice
parameters a = 0.586(3) nm, b = 0.392(7) nm, c = 1.009(5) nm and $\beta $ = 97.75(1)$^{\circ }$. Hence, the a and c - axis lattice parameters of Ni$%
_{0.05}$ZrTe$_{3}$ are larger than those of ZrTe$_{3}$. The substitution on
Zr site is unlikely due to the smaller size of Ni ion (r$_{Ni^{2+}}$ = 69 pm)
than Zr ion(r$_{Zr^{4+}}$ = 72 pm). This indicates that Ni is most likely
intercalated in the Van-der-Waals bonded crystallographic layers in ZrTe$%
_{3} $. The XRD pattern of single crystals (ig. 1(b) inset) reveals that
the crystal surface is normal to the c axis with the plate-shaped surface
parallel to the \emph{ab}-plane (fig. 1(c)).

\begin{figure}[tbp]
\centerline{\includegraphics[scale=1.0]{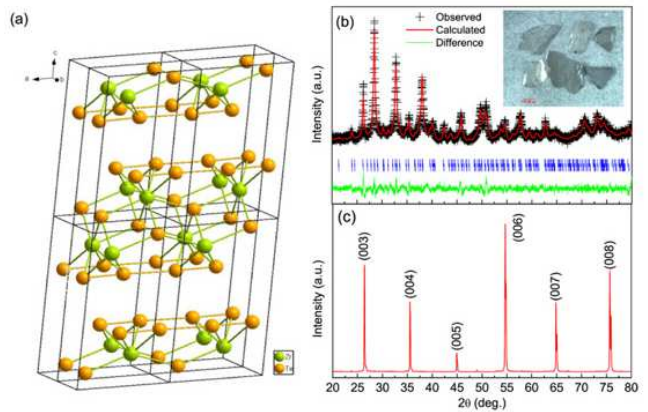}} \vspace*{-0.3cm}
\caption{(\textrm{a}) The structure of ZrTe$_{3}$ with green (orange)
symbols for Zr (Te) atoms. (\textrm{b}) Powder XRD pattern of Ni$_{0.05}$ZrTe$%
_{3}$, Inset: photo of typical single crystals of Ni$_{0.05}$ZrTe$_{3}$. (%
\textrm{c}) XRD result of Ni$_{0.05}$ZrTe$_{3}$ single crystal.}
\end{figure}

\begin{figure}[tbp]
\centerline{\includegraphics[scale=0.8]{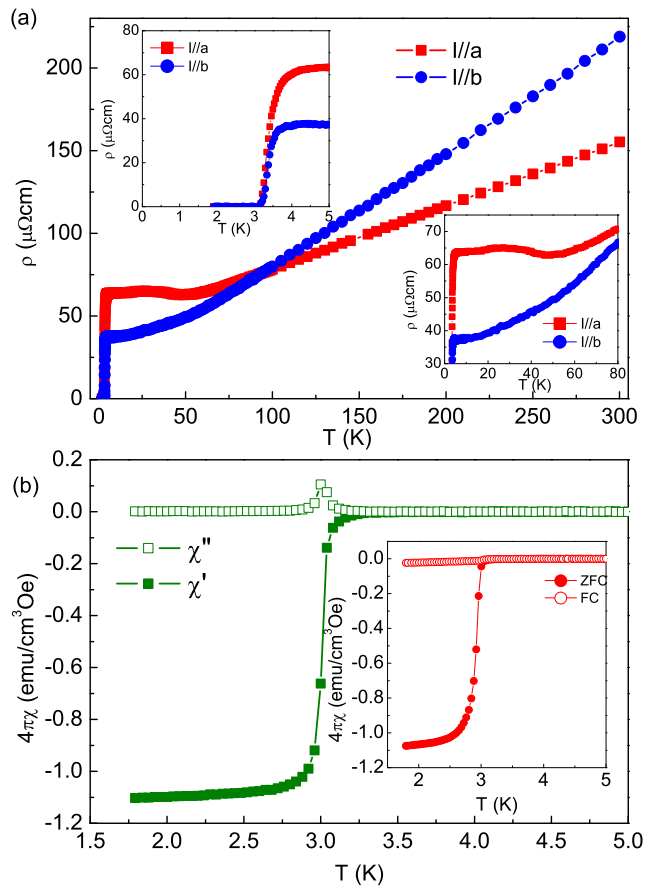}} \vspace*{-0.3cm}
\caption{(\textrm{a}) Temperature dependence of the resistivity $\protect%
\rho _{a}(T)$, $\protect\rho _{b}(T)$ of Ni$_{0.05}$ZrTe$_{3}$. Inset:
resistivity near $T_{c}$. (\textrm{b}) Temperature dependence of ac magnetic
susceptibility of Ni$_{0.05}$ZrTe$_{3}$. Inset: temperature dependence of dc
magnetic susceptibility with ZFC and FC.}
\end{figure}

Figure 2(a) shows the temperature dependence of resistivity at zero field
from 1.9 K to 300 K. Both $\rho _{a}$(T), and $\rho _{b}$(T) undergo a
relatively sharp superconducting transition at $T_{c,onset}$ = 3.74(3) K and
3.59(2) K, respectively (upper inset). It should be noted that the $\rho
_{b} $(T) is metallic in the normal state with no anomalies due to CDW and
with a residual resistivity ratio (RRR) of 6. The $\rho _{a}$(T) exhibits a
transition located at 41 K (lower inset fig. 2(a)). This can be ascribed to
CDW since it is similar to resistivity anomalies in ZrTe$_{3}$\cite%
{Takahashi}. However, the transition has been depressed from around 65 K to
41 K. Therefore, there is a coexistence of superconductivity and CDW in Ni$%
_{x}$ZrTe$_{3}$ similar to ZrTe$_{3}$ but with nearly double bulk $T_{c}$%
\cite{Nakajima}. The filamentary superconductivity and the CDW originate
from different sections of FS in ZrTe$_{3}$. From the difference in $T_{c}$,
$T_{CDW}$ in Ni$_{0.05}$ZrTe$_{3}$ and ZrTe$_{3}$ we can conclude that Ni
intercalation dramatically changes the FS related to the superconductivity
but has a minor effects on FS parts associated with the CDW transition.
Intercalation has a distinctively different effect when compared to
pressure. The reason could be that intercalation only modifies one part of
FS related to superconductivity whereas pressure increases
three-dimensionality of all parts.

Figure 2(b) shows the temperature dependence of the ac susceptibility of Ni$%
_{0.05}$ZrTe$_{3}$ single crystal with dimension of 2$\times $1.5$\times $%
0.24 mm$^{3}$ for H$\parallel $ab and H$\parallel $c. The single sharp peak
of 4$\pi \chi^{\prime\prime}$ accompanied by a very steep transition in 4$\pi \chi
^{\prime }$, indicates that the sample is rather homogeneous. The onset
temperature for both field direction is 3.1 K. This is slightly lower than
that obtained from resistivity measurement. The transition width $\Delta
T_{c}$ is 0.3 K. For H$\parallel $ab, the value of -4$\pi \chi ^{\prime }$
at 1.8 K is 110\% without any demagnetization factor correction, indicating
that the superconductivity is bulk. This is different from the mother
compound ZrTe$_{3}$ where superconductivity is undetectable in magnetization
due to its filamentary nature below $T_{c}$ = 2 K. Inset in fig. 2(b) shows
the dc magnetic susceptibility for field along $\widehat{a}$, $\widehat{c}$
axes with zero-field cooling (ZFC) and field cooling (FC). For ZFC curves,
the $T_{c,onset}$ and $\Delta T_{c}$ are consistent with the results of ac
susceptibility and the values of -4$\pi \chi $ at 1.8 K are nearly
identical. On the other hand, the volume fraction estimated from the FC
curve is about 2\% at 1.8 K for H$\parallel $c, similar to other
intercalated compounds such as (Pyridine)$_{1/2}$TaS$_{2}$\cite{Prober}, YbC$%
_{6}$\cite{Weller}, Ni$_{x}$TaS$_{2}$\cite{Li}. The small magnetization
values for FC is likely due to the complicated magnetic flux pinning effects
in the intercalated compounds\cite{Prober}.

\begin{figure}[tbp]
\centerline{\includegraphics[scale=0.45]{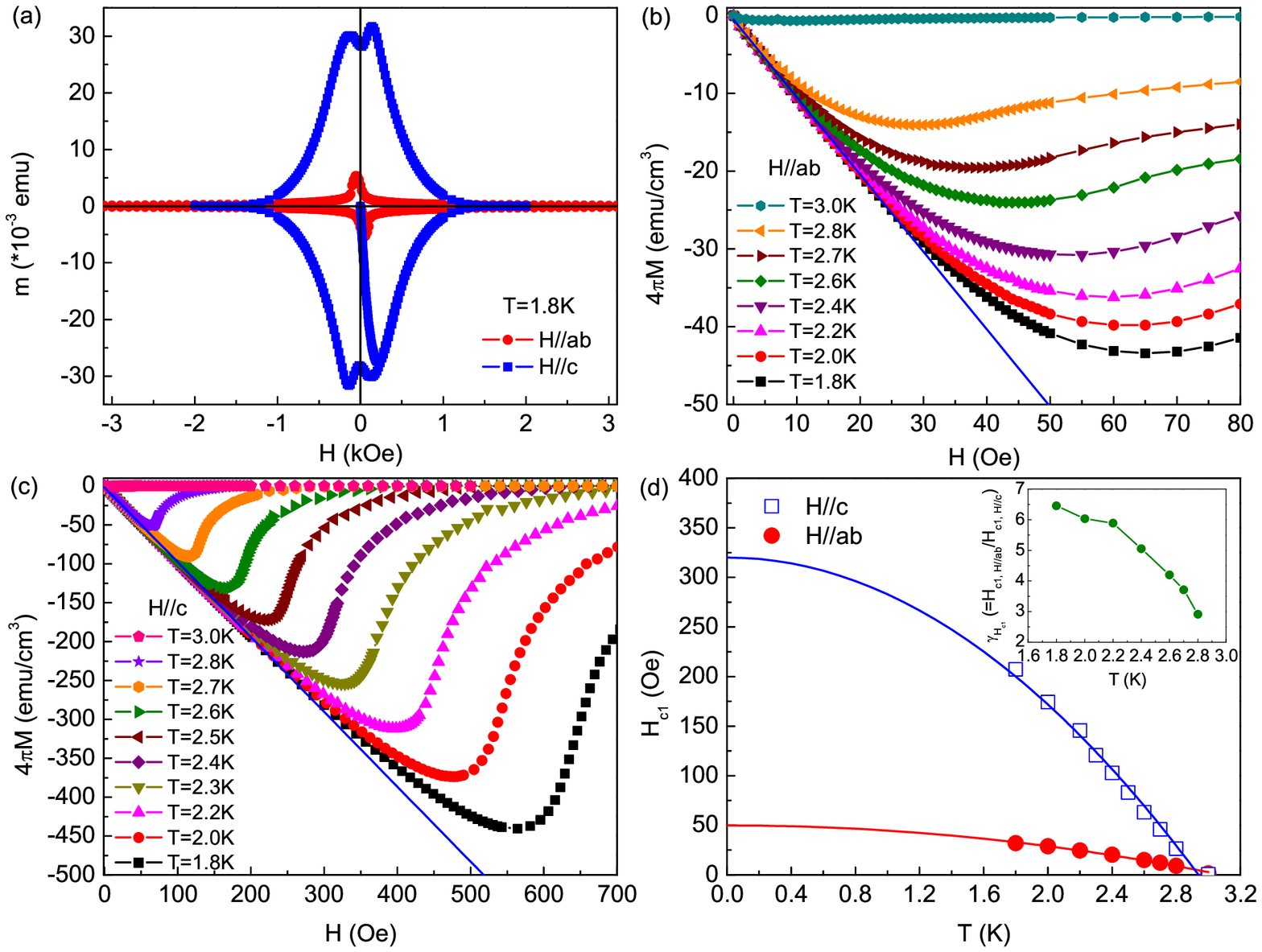}} \vspace*{-0.3cm}
\caption{(\textrm{a}) Magnetization hysteresis loops of Ni$_{0.05}$ZrTe$_{3}$
at T = 1.8 K for H$\Vert $ab and H$\Vert $c. (\textrm{b}) and (\textrm{c}) Low
field parts of M(H) at various temperature for H$\Vert $ab and H$\Vert $c
with demagnetization correct, respectively. The two solid red lines are the
\textquotedblleft Meissner line\textquotedblright\ as discussed in the text.
(\textrm{d}) Temperature dependence of $H_{c1}$ for H$\Vert $ab and H$\Vert $%
c. The dashed lines are the fitted lines using $H_{c1}$ = $H_{c1}$(0)(1-($T/T%
_{C}$)$^{2}$). Inset: the temperature dependence of anisotropy of H$_{c1}$, $%
\protect\gamma _{_{H_{c1}}}$ = $H_{c1,ab}$(T)$/$$H_{c1,c}(T)$.}
\end{figure}

The results of the dc magnetization versus field M(H) at various
temperatures for both directions are shown in fig. 3(a). The shape of the
M(H) curves points that Ni$_{0.05}$ZrTe$_{3}$ is a typical type-II
superconductor. Figures 3(b) and (c) show
the initial M(H) curves at the low field region. All curves clearly fall on
the same line and deviate from linearity for different temperatures. Linear
fits describe the Meissner shielding effects (\textquotedblleft Meissner
line\textquotedblright ) (figs. 3(b) and (c)) The obtained slope of the
linear fit up to 30 Oe at the lowest temperature of our measurement T = 1.8
K is -0.995(5)$\approx $ -1. This corresponds to -4$\pi $M = H for H$%
\parallel $ab where the demagnetization factor is negligible.

The value of $H_{c1}^{\ast }$ at which the field starts to penetrate into
the sample can be determined by examining the point of deviation from the
Meissner line on the initial slope of the magnetization curve. The $H%
_{c1}^{\ast }$ is not the same as the real lower critical field, $H_{c1}$,
due to the geometric effect. The $H_{c1}$ can be deduced from the first
penetration field $H_{c1}^{\ast }$, assuming that the magnetization M = -$H%
_{c1}$ when the first vortex enters into the sample. Thus H has been
rescaled to H = H$_{a}$-NM and $H_{c1}$ = $H_{c1}^{\ast }$/(1-N) where N is the
demagnetization factor and H$_{a}$ is the external field\cite{Fossheim}. We
estimate demagnetization factors 0.075 and 0.781 for H$\parallel $ab and H$%
\parallel $c by using $H_{c1}$ = $H_{c1}^{\ast }$/tanh($\sqrt{0.36b/a}$),
where a and b are width and thickness of a plate-like superconductor\cite%
{Brandt}. This is consistent with previous, i.e. the demagnetization factor
for H$\parallel $ab is negligible. However, demagnetization factor must be
considered for H$\parallel $c. The M(H) curve for H$\parallel $c with
demagnetization correction is shown in fig. 3(c) and the slope of the fitted
line in linear region is -0.963(1) $\approx $ -1. From the above and
constrained with the resolution limit of our magnetization measurement $%
\Delta m$= 3$\times $10$^{-5}$ emu we extract the temperature dependence of $H%
_{c1}$ for both field directions (fig. 3(d)). The $H_{c1}$(T) can be well
explained by $H_{c1}$(T) = $H_{c1}$(0)[1 - ($T/T_{c}$)$^{2}$]. The obtained $H%
_{c1,ab}$(0) and $H_{c1,c}$(0) are 50(1) and 321(8) Oe, respectively. On
the other hand, the anisotropy of $H_{c1}$, $\gamma _{_{H_{c1}}}$ = $H%
_{c1,H//c}$$/$$H_{c1,H//ab}$, increases with decreasing temperature and is
estimated about 6.4 at T = 1.8 K. This is much larger than in MX$_{2}$
compounds\cite{Li}.

\begin{figure}[tbp]
\centerline{\includegraphics[scale=0.45]{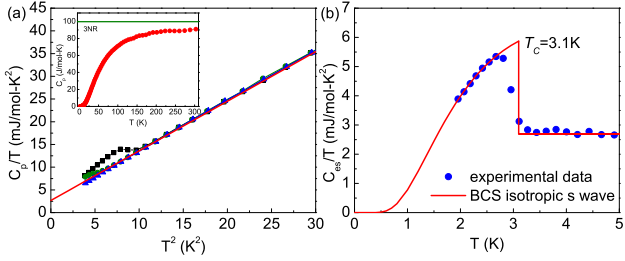}} \vspace*{-0.3cm}
\caption{(\textrm{a}) Low temperature specific heat behavior of Ni$_{0.05}$%
ZrTe$_{3}$ plotted as C$_{p}$/T vs T at H = 0, 1 and 50 kOe.
The solid line is a fit described in the text. Inset: temperature dependence
of $C_{p}$(T) from 1.95 K to 300 K at H = 0 kOe. (\textrm{b})
Temperature dependence of the electronic specific heat plotted as $C_{es}$/T
vs T at H = 0 kOe. The solid line shows fitted result of $C%
_{es}$/T assuming an isotropic s-wave BCS gap.}
\end{figure}

The main panel of fig. 4(a) shows the temperature dependence of the specific
heat of Ni$_{0.05}$ZrTe$_{3}$ below 5 K at H = 0, 10, and 50 kOe.\ A jump at 3
K can be clearly seen at H = 0, indicating bulk superconductivity of Ni$%
_{0.05}$ZrTe$_{3}$ sample. The $T_{c,onset}$ of 3.1 K determined from the
specific heat jump is consistent with that obtained from the susceptibility
and transport measurements. At H = 1 kOe, the superconducting anomaly is
shifted to lower temperatures. The normal state is recovered at H = 50 kOe. In
order to obtain the electronic specific heat coefficient $\gamma $ and Debye
temperatures $\Theta _{D}$, the low temperature specific heat at H = 50 kOe is
fitted using $C_{p}/T$ = $\gamma +\beta T^{2}$. The obtained parameters are $%
\gamma $ = 2.7(1) mJ/mol-K$^{2}$ and $\Theta _{D}$ = 192.4(1) K using $\Theta
_{D}$ = $(12\pi ^{4}NR/5\beta )^{1/3}$ where N = 4 is the number of atoms per
formula unit and R is the gas constant. The high-temperature specific heat
approaches the value of 3NR, in accordance with the Dulong-Petit law (fig.
4(a) inset). According to the McMillan formula for electron-phonon mediated
superconductivity\cite{McMillan}, the electron-phonon coupling constant $%
\lambda $ can be determined by

\begin{equation}
T_{c}=\frac{\Theta _{D}}{1.45}\exp [-\frac{1.04(1+\lambda )}{\lambda -\mu
^{\ast }(1+0.62\lambda )}],
\end{equation}

where $\mu ^{\ast }\approx $ 0.13 is the common value for Coulomb
pseudopotential. By using $T_{c}$ = 3.1 K and $\Theta _{D}$ = 192.4 K, we obtain
$\lambda $ = 0.63, indicating an intermediately coupled superconductor. The
electronic specific heat part $C_{es}$ obtained by subtracting phonon part
from the total specific heat is shown in fig. 4(b). From fits to $C_{es}$/$T$%
-$T$ below $T_{c}$ using the BCS formula for the electronic contribution to
the specific heat\cite{Bardeen}, the ratio of the gap and the critical
temperature is about, 2$\Delta $/$k_{B}T_{c}$ = 3.00(2). This is smaller than
the typical BCS value (3.53) in the weak-coupling limit\cite{Bardeen}. The
solid red line in fig. 4(b) shows the data simulation for an isotropic
s-wave BCS gap. The specific heat jump at $T_{c}$, $\Delta $C$_{es}$/$\gamma
T_{c}$ = 1.20(3), is somewhat smaller than the weak coupling value 1.43\cite%
{McMillan}. The calculated coupling strength from the heat capacity anomaly
could be influenced by the rather large fitting inaccuracy near T$_{c}$ and
somewhat broad superconducting transition width in $C/T$($T$).

Temperature dependent resistivity of $\rho _{b}$(T) of Ni$_{0.05}$ZrTe$_{3}$
below 5 K in magnetic fields for H$\Vert $a and H$\Vert $c is shown in figs.
5(a) and (b). With increasing magnetic fields, the resistivity transition
width becomes broader and the onset of superconductivity gradually shifts to
lower temperatures. At H = 5 kOe, for H$\Vert $c, the superconducting
transition can not be observed above 1.9K, whereas for H$\Vert $a, the
superconductivity above 1.9 K is nearly suppressed at H = 10 kOe.

\begin{figure}[tbp]
\centerline{\includegraphics[scale=0.8]{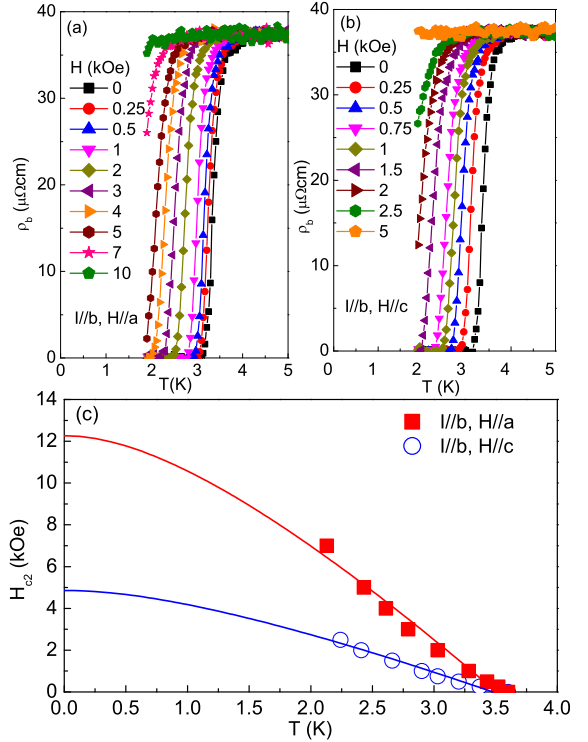}} \vspace*{-0.3cm}
\caption{Temperature dependence of the resistivity $\protect\rho _{b}(T)$ of
Ni$_{0.05}$ZrTe$_{3}$ for (\textrm{a}) H$\Vert $a and (\textrm{b}) H$\Vert $c
at the various magnetic fields. (\textrm{c}) Temperature dependence of the
upper critical field H$_{c2}$ for H$\Vert $a and H$\Vert $c. Solid lines
show the fitted result using WHH formula described in the text.}
\end{figure}

\begin{table*}[tbp]
\caption{Superconducting parameters of Ni$_{0.05}$ZrTe$_{3}$.}
\label{TableKey}\centering
\begin{tabular}{c|cccccc}
\hline\hline
Ni$_{0.05}$ZrTe$_{3}$ & $H_{c1}$(0) (Oe) & $H_{c2}$(0) (kOe) & $H_{c}$(0) (Oe)
& $\xi $(0) (nm) & $\lambda $(0) (nm) & $\kappa $(0) \\ \hline
H$\Vert $ab & 50(1) & 12.281(860) & 449(13) & 26(1) & 78(11) & 19.3(8) \\
H$\Vert $c & 321(8) & 4.862(340) &  & 10(1) & 1242(100) & 3.0(3) \\
\hline\hline
\end{tabular}%
\end{table*}

Figure 5(c) shows the upper critical field $H_{c2}(T)$ of Ni$_{0.05}$ZrTe$_{3}$
corresponding to the temperatures where the resistivity drops to 90\% of the
normal state resistivity $\rho _{n,b}(T,H)(T_{c,onset})$ at 5 K.\ Since
Pauli limiting field $H_{p}(0)$ = 1.84$T_{c}$ $>$ 55 kOe, the orbital effect
should be the dominant pair-breaking mechanism. According to the
conventional one-band Werthamer-Helfand-Hohenberg (WHH) theory, which
describes the orbital limited upper critical field of dirty type-II
superconductors\cite{Werthamer}, the $H_{c2}$ can be described by

\begin{equation}
\ln \frac{1}{t}=\psi (\frac{1}{2}+\frac{\overline{h}}{2t})-\psi (\frac{1}{2}),
\end{equation}

where $t$ = $T/T_{c}$, $\psi $ is a digamma function and

\begin{equation}
\overline{h}=\frac{4H_{c2}}{\pi ^{2}T_{c}(-dH_{c2}/dT)_{T=T_{c}}}.
\end{equation}

The $H_{c2}(T)$ fits are shown by solid and dotted lines in fig. 5(c). The
obtained $-dH_{c2}/dT|_{T=T_{c}}$ = 5.049(285) kOe, $T_{c}$ = 3.51(4) K for H$%
\Vert $a and $-dH_{c2}/dT|_{T=T_{c}}$ = 2.016(122) kOe, $T_{c}$ = 3.48(4) K for H%
$\Vert $c are consistent with the $T_{c,onset}$ ($\rho _{b}$ = 90\%$\rho_{n,b}$(T, H = 0) = 3.59(2) K). The estimated upper critical fields are $H_{c2}(0)$ = 12.281(860) kOe and 4.862(340) kOe for H$\Vert $a and H$\Vert $c.
From the $H_{c2}(0)$ zero-temperature coherence length $\xi (0)$ can be
estimated with Ginzburg-Landau formula $H_{c2,c}(0)$ = $\Phi _{0}/[2\pi \xi
_{ab}^{2}(0)]$, and $H_{c2,ab}(0)$ = $\Phi _{0}/[2\pi \xi _{ab}(0)\xi _{c}(0)]$
where $\Phi _{0}$ = 2.07$\times $10$^{-15}$ Wb. Based on the values of $%
H_{c1}(0)$ and $H_{c2}(0)$, the Ginzberg-Landau (GL) parameter $\kappa $ is
obtained from $H_{c2}(0)/H_{c1}(0)$ = $2\kappa ^{2}/(ln\kappa +0.08)$. And
thermodynamic critical field $H_{c}(0)$ can be obtained from $H_{c}(0)$ = $%
H_{c1,ab}(0)/[\sqrt{2}\kappa _{ab}(0)]$. The GL penetration length $\lambda
_{ab}$(0) and $\lambda _{c}$(0) can be evaluated using $\kappa _{c}$(0) = $%
\lambda _{ab}$(0)/$\xi _{ab}$(0), and $\kappa _{ab}$(0) = [$\lambda _{ab}$(0)$%
\lambda _{c}$(0)/$\xi _{ab}$(0)$\xi _{c}$(0)]$^{1/2}$\cite{Plakida}. All of
obtained parameters are listed in Table 1.

\section{Conclusion}

In summary, we have discovered the bulk superconductivity in Ni intercalated
ZrTe$_{3}$ and presented detailed characterization of superconducting state.
The bulk superconducting T$_{c}$ was nearly doubled when compared to
filamentary T$_{c}$ in parent material ZrTe$_{3}$. The normal state
electrical resistivity indicates that the CDW transition is somewhat
suppressed with Ni intercalation, coexisting with bulk superconductivity.
The Ni intercalation substantially changes the FS related to the
superconductivity but has a minor effects on FS parts associated with CDW.
This is rather different from the pressure effects. Our results suggest
intermediately coupled superconductivity and possible rich vortex physics.

\acknowledgments We thank John Warren for experimental support at Brookhaven
National Laboratory (BNL). This work was carried out at BNL, which is
operated for the U.S. Department of Energy by Brookhaven Science Associates
DE-Ac02-98CH10886.

*Present and permanent address: High Magnetic Field Laboratory, Chinese
Academy of Sciences, Hefei 230031, People's Republic of China

\end{document}